\documentclass[journal]{vgtc}                     

\onlineid{0}
\usepackage[table,svgnames,HTML]{xcolor}
\definecolor{NavyBlue}{RGB}{0,0,128}
\usepackage{multirow}
\usepackage{booktabs}
\usepackage{siunitx}
\usepackage{graphicx}
\usepackage{subcaption}

\vgtccategory{Research}

\vgtcpapertype{please specify}

\title{MisVisFix: An Interactive Dashboard for Detecting, Explaining, and Correcting Misleading Visualizations using Large Language Models}

\author{
  \authororcid{Amit Kumar Das}{0000-0002-2600-8321},
  \authororcid{Klaus Mueller}{0000-0002-0996-8590}
}

  \authorfooter{
    \item
    Amit Kumar Das and Klaus Mueller are with the Computer Science Department, Stony Brook University, USA (E-mails: amitkumar.das@stonybrook.edu, mueller@cs.stonybrook.edu).

  }

\abstract{%
Misleading visualizations pose a significant challenge to accurate data interpretation. While recent research has explored the use of Large Language Models (LLMs) for detecting such misinformation, practical tools that also support explanation and correction remain limited. We present MisVisFix, an interactive dashboard that leverages both Claude and GPT models to support the full workflow of detecting, explaining, and correcting misleading visualizations.
MisVisFix correctly identifies 96\% of visualization issues and addresses all 74 known visualization misinformation types, classifying them as major, minor, or potential concerns. It provides detailed explanations, actionable suggestions, and automatically generates corrected charts. An interactive chat interface allows users to ask about specific chart elements or request modifications.
The dashboard adapts to newly emerging misinformation strategies through targeted user interactions. User studies with visualization experts and developers of fact-checking tools show that MisVisFix accurately identifies issues and offers useful suggestions for improvement. By transforming LLM-based detection into an accessible, interactive platform, MisVisFix advances visualization literacy and supports more trustworthy data communication.

}


\teaser{
  \centering
  \includegraphics[width=\linewidth, alt={MisVisFix dashboard design showing key interface components}]{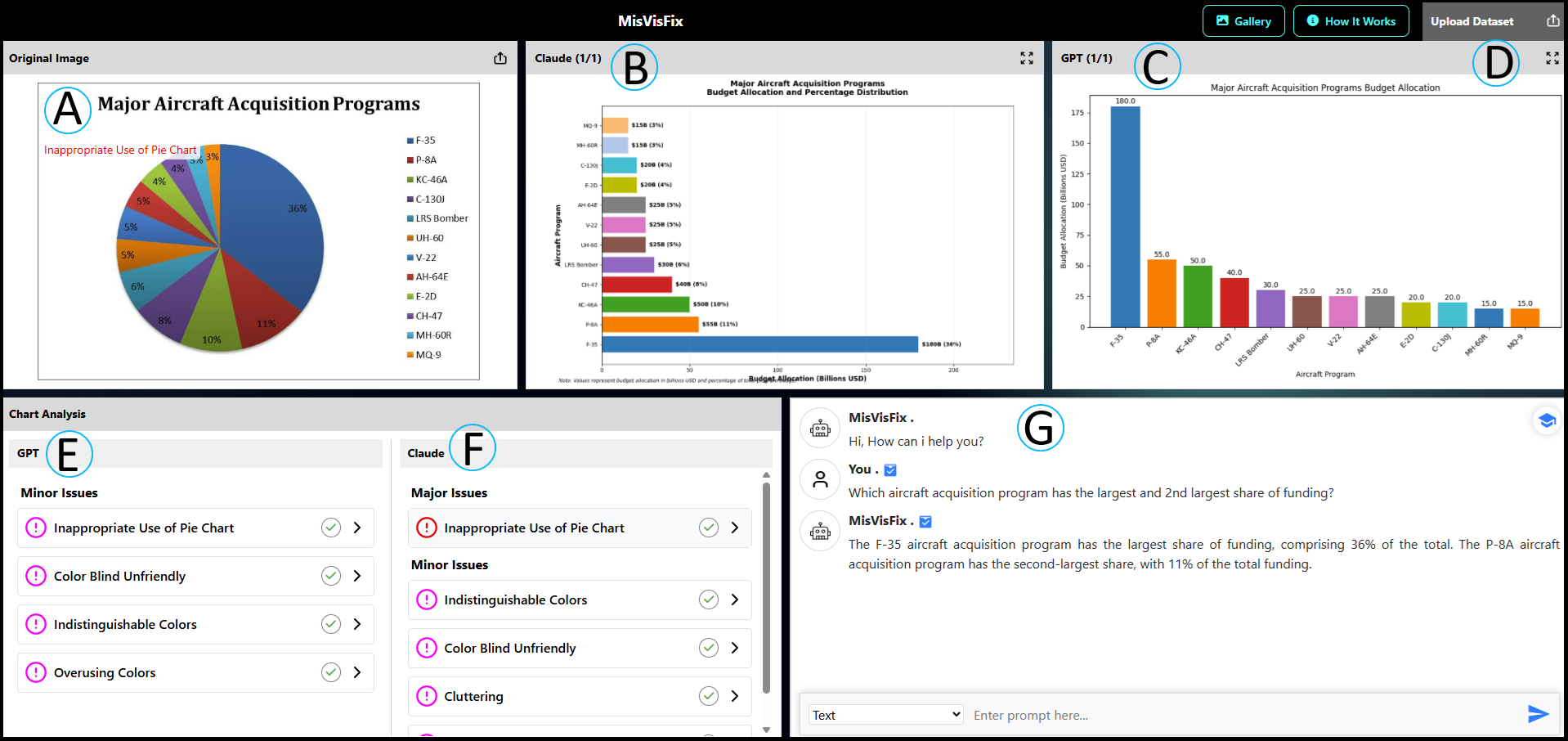}
  \caption{%
  	MisVisFix dashboard. Panel A displays the original misleading visualization with interactive issue localization, where hovering over identified issues highlights the corresponding problematic regions directly on the chart. Panels B and C show corrected versions generated by Claude and GPT, enabling side-by-side comparison. Panel D allows users to upload the original dataset to improve accuracy when data extraction fails. Panels E and F list detected issues from GPT and Claude, categorized by severity. Panel G contains the interactive chat, where users can request modifications and view updated visualizations.
  }
  \label{fig:teaser}
}

\graphicspath{{figs/}{figures/}{pictures/}{images/}{./}} 

\usepackage{tabu}                      
\usepackage{booktabs}                  
\usepackage{lipsum}                    
\usepackage{mwe}                       

\usepackage{mathptmx}                  

\begin{document}


\maketitle
\section{Introduction}

Data visualizations are powerful tools for communicating complex information clearly and efficiently, enabling users to comprehend patterns, trends, and relationships within datasets \cite{IN1,IN2}. However, misleading visualizations, whether created intentionally to deceive or inadvertently through poor design choices, can significantly distort a viewer's perception of the underlying data \cite{IN3,IN4,IN5}. These misleading representations span various techniques, from obvious manipulations like truncated axes to subtle distortions such as inappropriate encoding choices or unjustified data filtering \cite{NW105,NW106,NW107,NW108}. As visualizations play an increasingly central role in data-driven decision-making, identifying and correcting misleading visualizations has become increasingly critical \cite{IN6,IN7,NW109,NW110}.

\par Recent taxonomic work by Lo et al. \cite{IN8} systematically categorized 74 misleading issues that can arise in visualizations. This comprehensive framework has established a foundation for understanding how visualizations can mislead viewers. However, detecting these issues requires substantial expertise in visualization design principles and critical data thinking skills, which many consumers of data visualizations may lack \cite{IN9,IN10}. This expertise gap creates a pressing need for automated systems that can assist in identifying and correcting potentially misleading visualizations.

\par The emergence of multimodal Large Language Models (LLMs) with advanced vision capabilities presents new opportunities for addressing this challenge \cite{IN11,IN12}. Recent research has shown that models like GPT-4V and Claude 3 can interpret complex visual information and reason about visualization design issues \cite{IN13,IN14}. In particular, Alexander et al. \cite{IN15} explored the potential of GPT-4 in detecting misleading visualizations, concluding that these models can identify misleading elements with moderate accuracy when provided with appropriate guidance. Similarly, Lo et al. \cite{IN16} evaluated multiple LLMs for detecting misleading visualizations, finding significant potential but noting challenges in scaling detection capabilities across diverse issue types.

\par While existing research demonstrates the potential of LLMs for visualization analysis, several key limitations remain \cite{NW111,NW112,NW113}. First, most current approaches focus exclusively on detection without addressing the correction of problematic visualizations \cite{NW114,NW115,NW116}. Second, many systems target only a subset of misleading techniques rather than comprehensively addressing the complete taxonomy of potential issues \cite{NW117,NW118}. Third, limited work has been conducted on creating interactive, user-facing systems that effectively communicate identified issues and recommend improvements \cite{NW119,NW120}. These limitations present critical research opportunities for developing integrated systems that leverage the capabilities of LLMs across the complete pipeline of misleading visualization analysis.

\par Building on these identified gaps, we aim to address the following two key questions:

\begin{itemize} 
    \item \textbf{\textit{RQ1:}} 
    How can multimodal LLMs be effectively leveraged to detect and explain the full spectrum of misleading visualization techniques identified in existing taxonomies?
    \item \textbf{\textit{RQ2:}} To what extent can an interactive system facilitate both identification and correction of misleading visualizations, bridging the gap from detection and visualization best practices?

\end{itemize}

\par To address these research questions, we present MisVisFix, an interactive dashboard for detecting, explaining, and correcting misleading visualizations. Our work makes five primary contributions:

\begin{itemize} 
    \item
    We developed and implemented MisVisFix, an end-to-end interactive dashboard that integrates LLM capabilities to detect, explain, and generate corrections for misleading visualizations.

    \item
   We conducted a comprehensive evaluation of multimodal LLM performance across all 74 categories of visualization misinformation, demonstrating that our approach achieves an F1 score of 0.96 in issue detection.

    \item
   We created novel techniques for visually annotating misleading elements directly on visualizations with precise x-y positioning, enabling an intuitive understanding of problematic areas.

    \item
   We designed and implemented an interactive learning mechanism that continuously improves the system based on user feedback, allowing MisVisFix to adapt. 
   \item
    We validated the system's effectiveness through rigorous user evaluation with visualization experts and developers of fact-checking tools, confirming MisVisFix's accuracy, usefulness, and applicability in both professional and educational contexts.

\end{itemize}


Our paper is organized as follows. Section 2 presents related work. Section 3 describe the system design. Section 4 offers a system evaluation. Sections 5 and 6 end the paper with a discussion and conclusions. 

\section{Related Work}
Our research intersects several fields, including misleading visualization detection, visualization linters, chart analysis, and applications of large language models (LLMs) in visualization understanding. We review key developments in these areas to position our contributions within the broader research landscape

\subsection{Misleading Visualizations}
The study of misleading visualizations dates back decades, with foundational work by Huff \cite{RW1} and Tufte \cite{IN2} establishing core principles of graphical integrity. Huff's seminal text "How to Lie with Statistics" first exposed common deceptive practices in data representation, while Tufte introduced the concepts of "chartjunk" and "lie factor" to quantify visualization distortion. These early frameworks laid the groundwork for understanding how visual representations can manipulate viewer perception.

\par Recent empirical studies have advanced our understanding of how specific design choices impact interpretation. Pandey et al. \cite{IN3} systematically evaluated how visualization manipulations affect a viewer's perceptions, demonstrating that even subtle design choices like truncated axes significantly alter data interpretation. Correll et al. \cite{IN4}  examined y-axis truncation, revealing that this common practice systematically distorts the perception of data magnitudes. Additional research by Lee et al. \cite{IN6} and Lisnic et al. \cite{IN7} has explored how misinformation spreads through data visualizations in social media, emphasizing the societal impact of misleading charts in public discourse.

\par The most comprehensive taxonomy of misleading visualization practices was developed by Lo et al. \cite{IN8}, who identified 74 distinct types of visualization issues based on an analysis of over 1,000 real-world examples. This taxonomy spans both structural issues (e.g., truncated axes, inappropriate color schemes) and context-related problems (e.g., cherry-picked data, misrepresentation of statistical findings). Our research builds directly on this taxonomy, using it as the foundation for our detection framework.

\subsection{Visualization Linters}
Visualization linters—tools that algorithmically identify potential issues in visualizations—represent a significant advancement in automated quality assessment. These systems typically integrate with visualization creation libraries to detect issues during the authoring process. McNutt and Kindlmann \cite{RW8} pioneered this approach with a linter for matplotlib that implemented rules derived from visualization design principles. Chen et al. \cite{RW9} subsequently developed VizLinter, which analyzes visualization specifications created with Vega-Lite against established best practices using Answer Set Programming. Similar specialized linters have been developed for geographic visualizations, with Lei et al. \cite{RW10} creating GeoLinter for choropleth maps.


\par These linting systems offer valuable support but have key limitations: they operate during visualization creation rather than analyzing existing charts, focus on structural elements rather than contextual issues, and require access to underlying code rather than working with bitmap images. Our work addresses these gaps by analyzing bitmap visualizations and detecting both structural and contextual issues.

\subsection{Chart Analysis}
Computer vision approaches to chart analysis have made significant progress in recent years. Early work by Savva et al. \cite{RW11} established methods for extracting data from chart images through reverse engineering techniques. Poco and Heer \cite{RW12} built on this foundation with improved methods for recovering visual encodings from chart images. These reverse engineering approaches enable subsequent analysis of the extracted data and specifications.

\par Recently, research has expanded to chart question-answering tasks that require a deeper understanding of visualization semantics. Kahou et al. \cite{RW13} created the FigureQA dataset with over 100,000 chart images and corresponding question-answer pairs. Kafle et al. \cite{RW14} developed DVQA, a benchmark dataset focused on more complex reasoning about visualizations. These benchmarks have driven progress in systems that can answer natural language questions about chart content.

\par While these approaches demonstrate increasing capabilities in extracting information from charts, they primarily focus on understanding chart content rather than evaluating chart quality or identifying misleading elements. Our work extends these capabilities to critical assessment of visualization design and content.

\subsection{LLMs for Chart Understanding and Criticism}
The emergence of large language models with multimodal capabilities has created new opportunities for chart analysis. Recent research has demonstrated that these models can extract data from charts, answer questions about chart content, and even reason about chart design choices. Masry et al. \cite{RW15} developed UniChart, a vision-language model specifically trained for chart comprehension tasks. Do et al. \cite{IN13} investigated prompt engineering strategies for chart question answering with general-purpose multimodal LLMs, finding that well-designed prompts enable strong performance.

\par Most relevant to our work, Alexander et al. \cite{IN15} explored the capabilities of GPT-4 models in detecting misleading visualizations. Their study evaluated three variants of GPT-4 (4V, 4o, and 4o mini) across a dataset of tweet-visualization pairs containing various "misleaders." Their experiments incorporated different guidance levels, finding that GPT-4 models could detect misleading elements with moderate accuracy without prior training. Performance improved significantly when models were provided with definitions of potential issues. Lo and Qu \cite{IN16} comprehensively evaluated multiple multimodal LLMs in detecting misleading visualizations, testing nine distinct prompting strategies across three experiments. They found that the Chain of Thought prompting strategy was the most effective, but noted challenges in scaling detection to cover the full spectrum of potential issues. Similarly, Bendeck and Stasko \cite{MO5} evaluated GPT-4's visualization literacy capabilities, including its ability to identify deceptive and misleading visualizations as one of four key assessment dimensions, demonstrating moderate success in detecting visual deception techniques.

\par These studies demonstrate the potential of LLMs in detecting misleading visualizations but highlight several limitations. First, most existing work focuses on detection without addressing correction. Second, current approaches often struggle to scale beyond a limited set of issue types. Third, research has primarily evaluated model performance in controlled settings rather than developing complete, user-facing systems. Our work addresses these limitations by developing an interactive system that spans detection, explanation, and correction while 
addressing the full taxonomy of misleading visualization techniques.








\section{MisVisFix System Architecture}
The MisVisFix system employs a comprehensive architecture designed to facilitate the detection, explanation, and correction of misleading visualizations. We adopt a modular pipeline approach that decouples the distinct processing stages while maintaining integration across the entire system. This section provides a detailed description of the system architecture, core components, and implementation details.

\subsection{System Overview}
MisVisFix integrates multiple computational components and user interface elements to create a coherent workflow for visualization analysis. Fig. \ref{fig:teaser} presents a high-level overview of the system architecture, illustrating the primary components and their interactions. The pipeline begins with visualization input, processes the image through multiple analytical stages, and culminates in presenting results and corrected visualizations to the user.

\par The system architecture consists of four principal components: (1) the input processing module, (2) the visualization analysis engine, (3) the visualization correction and generation module, and (4) the interactive user interface. These components operate in sequence while facilitating iterative refinement through user interaction. The following sections detail each component's functionality and implementation.

\subsection{LLM Selection and Evaluation}
Our model selection process was based on direct testing of multiple LLMs for visualization analysis tasks. We conducted pilot studies to determine which models performed best to extract data from graphs and answer visualization-based questions.

\par We tested GPT-4.5, Claude-3.7, and GEMINI-2.0 for our evaluation. We chose these models because they represent the latest versions from leading LLM providers with multimodal capabilities. We considered other multimodal models, such as DeepSeek-V3, but limitations in their API capabilities (such as lack of support for image uploads through the API) prevented us from including them in our tests.

\par Previous research by Bendeck and Stasko \cite{MO5} and Hong et al. \cite{MO6} showed that older models like GPT-4.0 and Gemini 1.5 performed below human standards when answering chart questions. For our testing, we used the VLAT (Visualization Literacy Assessment Test) \cite{MO26}, a widely recognized assessment tool for visualization comprehension.

\par Instead of using the generic prompts from previous studies, we developed guided prompts that followed a three-step process: First, we asked the LLMs to extract data from the graphs. Then, we asked them to sort the extracted data. Finally, we asked them to answer questions based on this extracted data. Our results showed that significant improvement could be achieved with this step-wise approach. 

\par Table \ref{tab:1} shows the performance of the three LLMs on the original VLAT questions compared to the human baseline from Lee et al. \cite{MO26} and GPT-4.0 results reported by Bendeck and Stasko \cite{MO5}. We present both raw scores (out of 53 questions) and scores calculated using the VLAT scoring scheme.

\begin{table}[htbp]
\centering
\caption{Comparing the results we obtained with our 3-step prompting technique and existing methods, based on the full VLAT data set. Green text indicates performance improvements over the human baseline, while red text indicates performance below the baseline. Bold values highlight the best-performing model scores.}
\begin{tabular}{lcc>{\raggedleft\arraybackslash}p{1.5cm}}
\toprule
\textbf{Model} & \begin{tabular}[c]{@{}c@{}}\textbf{Mean Raw}\\\textbf{Score}\end{tabular} & \begin{tabular}[c]{@{}c@{}}\textbf{VLAT}\\\textbf{Score}\end{tabular} & \textbf{vs. Human} \\
\midrule
Human \cite{MO26} & 33.74 & 28.82 & baseline \\
GPT-4.0 \cite{MO5} & 29.33 & 19.67 & \textcolor{red}{\textbf{-31.7\%}} \\
GPT-4.5-preview & 42.00 & 34.33 & \textcolor{green!60!black}{\textbf{+19.1\%}} \\
GEMINI-2.0-pro & 40.00 & 33.50 & \textcolor{green!60!black}{\textbf{+16.2\%}} \\
Claude-3.7-sonnet & \textbf{51.00} & \textbf{50.17} & \textcolor{green!60!black}{\textbf{+74.1\%}} \\
\bottomrule
\end{tabular}

\label{tab:1}
\end{table}

\par All three models exceeded human baseline performance, but Claude-3.7-sonnet showed the highest accuracy, exceeding the human baseline by 74.1\%. This superior performance in data extraction made Claude-3.7 our primary choice for extracting data from visualizations in the MisVisFix dashboard.

\par While VLAT tests focus on data extraction and comprehension, these capabilities directly support misleading visualization analysis. Accurate data extraction enables comparison between chart content and underlying data to detect discrepancies. Chart comprehension skills help identify when visual elements misrepresent the data. Claude-3.7 and GPT-4.5 performed at comparably high levels, with GEMINI-2.0 also performing well—though slightly below GPT-4.5—leading us to focus on the two top-performing models.

\par To leverage the complementary strengths of both models, we implemented a dual-model approach in MisVisFix:

\begin{itemize} 
    \item Claude-3.7 handles the primary data extraction from charts due to its superior extraction accuracy.
    \item Both Claude-3.7 and GPT-4.5 contribute to issue detection and correction, offering users multiple perspectives on potential visualization problems.
\end{itemize}

\par This evidence-based approach to model selection ensures MisVisFix delivers reliable results across the entire pipeline from data extraction to visualization correction. The supplementary materials include all prompts, test procedures, and detailed performance results.

\subsection{Input Processing Module}
The input processing module serves as the entry point for visualizations into the system. This module accepts bitmap image uploads from users and performs preliminary processing to prepare the visualization for subsequent analysis. The preprocessing pipeline includes image normalization, data extraction, and initial chart-type detection.

\par To ensure consistent input quality, we implement standard preprocessing techniques for image normalization, including resizing, color space normalization, and artifact removal. We used Claude-3.7 for data extraction, which included titles, axis labels, legends, and annotations. Chart-type detection utilizes a multimodal LLM approach, where Claude 3.7 analyzes the image to classify the visualization.

\par The input module also handles metadata extraction, such as image dimensions, resolution, and file format. This information is preserved alongside the processed image for use in later pipeline stages, particularly for accurate placement of annotations when highlighting issues.

\subsection{Visualization Analysis Engine}
The visualization analysis engine constitutes the core analytical component of MisVisFix, responsible for detecting potential issues in the input visualization. This engine employs a multi-stage approach to comprehensive visualization analysis.
\subsubsection{Issue Detection Framework}

\par Following data extraction, the dashboard performs comprehensive issue detection across all 74 categories from Lo et al.'s taxonomy \cite{IN8}. We employ a structured detection framework that organizes issues into three tiers of severity:

\begin{itemize} 
    \item  \textbf{Major Issues:} Problems that substantially distort data perception or introduce significant misinterpretation risk (e.g., truncated axes, misleading color encodings, fabricated data).
    \item  \textbf{Minor Issues:} Problems that affect readability or clarity but do not fundamentally alter data interpretation (e.g., missing titles, inconsistent label formatting).
    \item  \textbf{Potential Issues:} Elements that may present problems depending on context or audience (e.g., specialized encoding choices, technical terminology).    

\end{itemize}

\begin{figure}[!thb]
    \centering
    
    \captionsetup[subfigure]{font=normalsize,labelformat=parens}
    
    \begin{subfigure}{\linewidth}
        \centering
        \includegraphics[width=0.99\linewidth, alt={First comparison}]{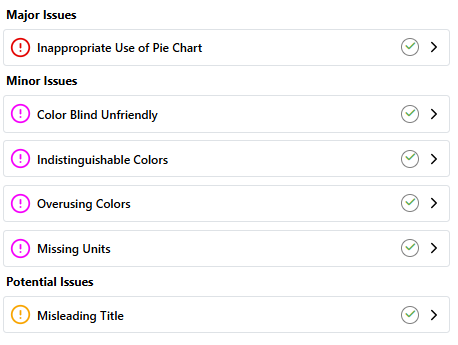}
        \caption{Dashboard view showing detected issues categorized by severity level}
        \label{fig:fig_issues_a}
    \end{subfigure}
    
    \vspace{0.1cm}
    
    \begin{subfigure}{\linewidth}
        \centering
        \includegraphics[width=0.99\linewidth, alt={Second comparison}]{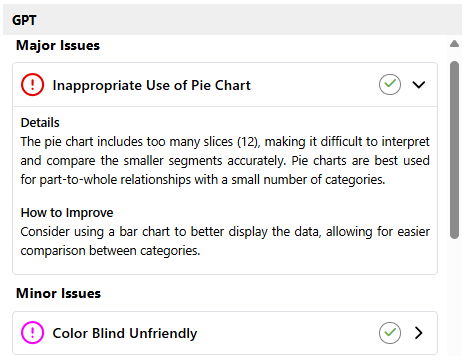}
        \caption{Expanded view of a selected issue showing detailed explanation and improvement recommendations}
        \label{fig:fig_issues_b}
    \end{subfigure}

    \caption{Example output from the MisVisFix issue analysis interface for a specific misleading visualization}
    \label{fig:fig_issues}
\end{figure}

Our correction module provides users with both issue detection and detailed explanations. Fig. \ref{fig:fig_issues} illustrates the interactive issue analysis interface of MisVisFix.

\par As shown in Fig. \ref{fig:fig_issues_a}, users can view all detected issues organized by severity categories (major, minor, potential). When users click on a specific issue, as demonstrated in Fig. \ref{fig:fig_issues_b}, the dashboard presents a detailed explanation of why this particular visualization element is problematic, along with specific recommendations for improvement. The green checkmarks indicate issues that have been successfully addressed in the generated corrected visualizations.
\par This interactive approach helps users understand both what visualization problems exist and how they can be fixed, creating a complete learning loop that enhances visualization literacy while providing practical solutions.

\par We utilize GPT-4.5 and Claude 3.7 for issue detection, each with specialized prompts designed to maximize detection accuracy. Our issue detection approach incorporates Chain of Thought techniques that guide the models through a structured reasoning process.

\begin{itemize} 
    \item \textbf{Comprehensive Issue Catalog:} We provide the models with a systematic listing of all 74 potential visualization issues identified in the literature \cite{IN8}. This ensures the models check for the complete range of possible problems.
    \item \textbf{Focused Analysis Request:} We direct the models to return only the names of identified issues without additional text, enabling efficient processing of results.
    \item \textbf{Detailed Issue Explanation:} After issues are identified, we prompt the models to provide detailed explanations for each issue, describing why they could be misleading or incorrect in the context of the specific visualization.
    \item \textbf{Issue Categorization:} The models organize identified issues by severity (major, minor, potential), helping users prioritize which problems to address first.

\end{itemize}

\par This structured prompting approach enables consistent analysis across visualization types while maximizing detection accuracy.

Our Chain of Thought prompting strategy guides LLMs through a systematic analysis process. The prompt instructs the model to (1) identify chart type and elements, (2) examine axis properties, (3) analyze color usage, (4) review data representation, and (5) list detected issues by severity category. 


\subsubsection{Issue Localization Technique}

\par Beyond detecting issues, MisVisFix provides precise location information. To help users locate issues (major, minor, and potential) in the original uploaded graph, we developed a technique that identifies the exact position of each problem area on the chart. As shown in Fig. \ref{fig:teaser} Panel (A), this allows direct annotation on the original graph. When users hover over an issue in the dashboard panels, the dashboard highlights the corresponding region on the chart where the problem exists. For example, if "Missing Title" is identified, hovering over this issue in the panel highlights the area at the top of the chart where a title should appear. This interactive approach helps users understand not just what issues exist, but exactly where they occur in the graph. The tooltip feature creates an immediate visual connection between abstract problems and their concrete manifestations in the chart.

\par Our issue localization technique maps each detected issue to precise coordinates on the visualization. We prompt the LLM with the visualization image and a list of detected issues. The model then analyzes the image and returns `top\_gap' and `left\_gap' percentage-based coordinates for each issue. For issues like `Truncated Axis,' the model pinpoints the specific axis region, while for `Color Blind Unfriendly' issues, it identifies relevant color-encoded elements. These coordinates create responsive highlights that work across different display sizes. When users hover over an issue in the dashboard panels, the system uses these stored coordinates to highlight the corresponding region on the chart, connecting abstract problems to their visual location.


\subsection{Visualization Correction and Generation}
A distinguishing feature of MisVisFix is its ability to identify issues and generate corrected visualizations. The correction module takes the original visualization, the extracted data, and the identified issues as input and then produces a corrected visualization that addresses the detected problems.

\par
\begin{figure}[!thb]
    \centering
    \includegraphics[width=0.48\textwidth, height=3cm, alt={Comparison}]{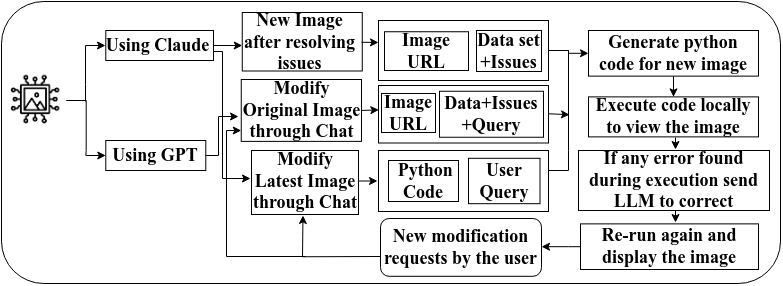}
    \caption{Visualization generation workflow using Claude and GPT. The system analyzes uploaded charts, generates Python code for improved versions, and enables iterative refinement through interactive chat until users are satisfied with the results.}
    \label{fig:fig_9}
\end{figure}

 The correction process employs a code generation approach, where the LLM generates Python code using matplotlib, seaborn, or other visualization libraries to recreate the visualization without the identified issues. Fig. \ref{fig:fig_9} illustrates the overall working procedure of graph generation using Claude and GPT, showing how the system initially generates graphs and then iteratively refines them based on user input.

\par We encountered several technical challenges during the implementation of the correction module, particularly related to code execution and image generation. Initial attempts using DALL-E for chart generation proved unsuccessful, as these models are optimized for creative image generation rather than precise data visualization. We addressed these challenges through a series of refinements:

\begin{itemize} 
    \item Pre-installing a consistent set of visualization libraries in a controlled execution environment.
    \item Implementing error handling mechanisms that send Python code and corresponding errors back to the LLM for correction.
    \item Converting generated plots to base64 encoded images for consistent rendering and implementing Firebase storage for generated images that exceed token limits when sent directly to LLMs. 

\end{itemize}

\par
\begin{figure}[!thb]
     \centering
     \includegraphics[width=0.5\textwidth, alt={Comparison}]{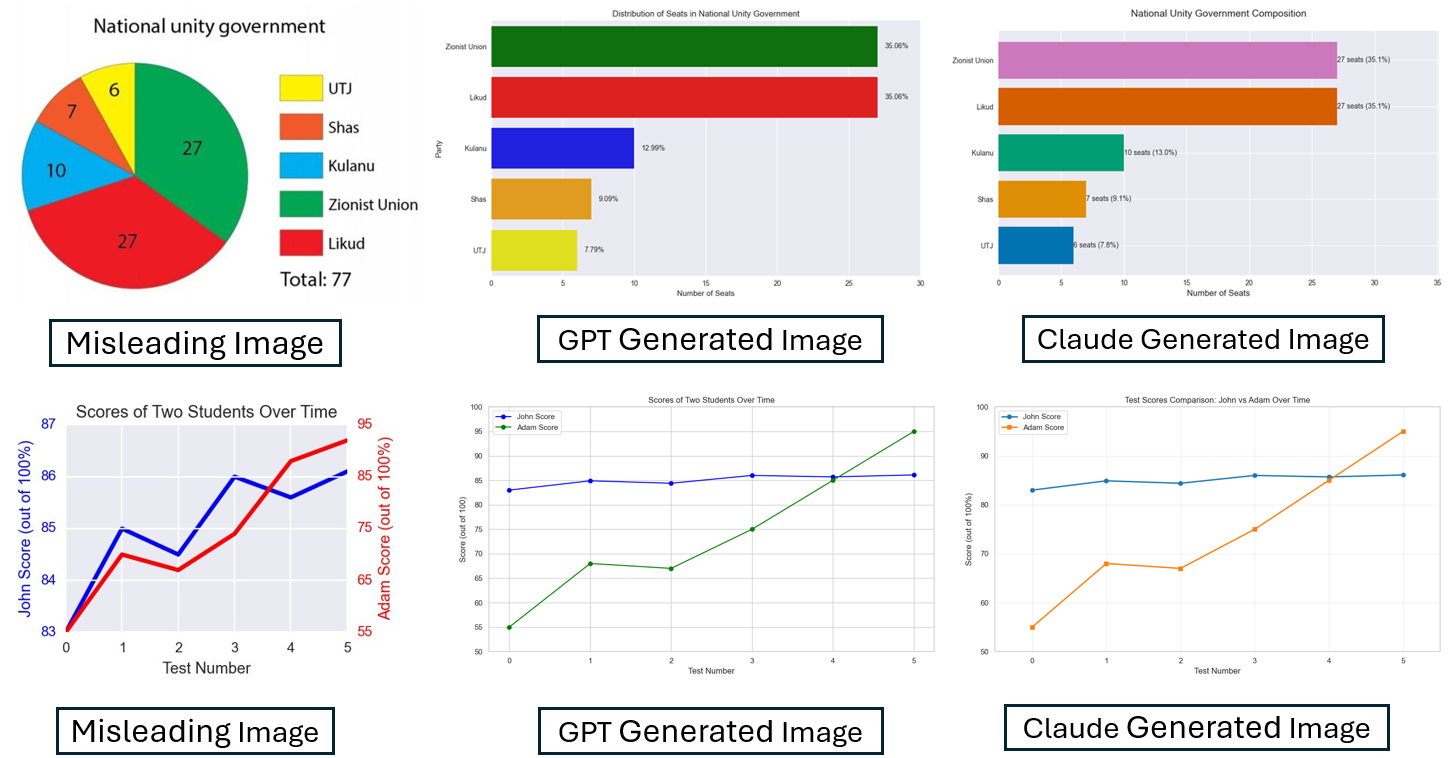}
     \caption{Comparison of original misleading visualizations and MisVisFix-generated corrections. Top: Pie chart with too many similar-colored segments converted to sorted bar chart after detecting inappropriate chart type, indistinguishable colors, and missing units. Bottom: Line chart with dual y-axes causing data misrepresentation replaced with single-axis line charts after identifying dual axis issues, data magnitude differences, and misrepresentation problems. Both examples show conversion to clearer chart formats that reduce misinterpretation and improve data comprehension.}
     \label{fig:fig_5}
 \end{figure}

 This robust approach ensures the reliable generation of corrected visualizations that maintain visual similarity to the original while addressing the identified issues. Fig. \ref{fig:fig_5} presents examples of original misleading visualizations alongside their MisVisFix-generated corrections, demonstrating the system's ability to maintain design integrity while correcting problematic elements.

\par Our correction approach prioritizes standard visualization types like bar charts, line charts, and scatterplots that have less potential to mislead viewers. As shown in Fig. \ref{fig:fig_5}, when the system detects fundamentally problematic visualization choices, such as pie charts with too many segments or maps that distort data through geographic projection, it may completely replace the original design with a more appropriate chart type rather than attempting to fix the original format. This transformation ensures that the underlying data relationships remain intact while removing elements that could lead to misinterpretation.

\par Our correction module uses a template-based code generation process that adapts to specific chart types and issues. For example, when fixing truncated axes, the system generates Python code using Matplotlib with corrected axis limits. The process follows three steps: (1) extract data from the original visualization, (2) generate visualization code with fixes for the identified issues, and (3) apply styling to match the original.
\subsection{Interactive User Interface}
The MisVisFix dashboard provides an intuitive interface that integrates all system components into a cohesive user experience. The interface is shown in Fig. \ref{fig:teaser} and centers around multiple visualization panels: Panel A displays the original misleading visualization uploaded by the user. Panels B and C show corrected versions of the visualization—Panel B contains the version generated by Claude, while Panel C shows the version created by GPT. This side-by-side presentation allows users to compare the original problematic visualization with two alternative corrections, facilitating critical evaluation of different approaches.

\par Panel D offers a dataset upload functionality, addressing cases where LLMs fail to extract data correctly from the visualization. This feature allows users to provide the original dataset directly, enabling the system to fine-tune its analysis and generate more accurate corrected visualizations based on the actual data rather than extracted approximations.

\par Panels E and F display the misleading issues identified by GPT and Claude, respectively. Each panel organizes issues by severity level (major, minor, potential) and provides detailed descriptions of each detected problem. While both models often identify similar core issues, they sometimes differ in detecting subtle problems or explaining approaches, providing complementary perspectives.

\par Panel G houses the interactive chat window, where users can ask questions about the visualization and request specific modifications. For example, if the visualization uses a green color scheme, users can request, ``Can you make it blue?" The system processes these requests and generates updated visualizations incorporating the requested changes. All versions created through this interactive process remain accessible, allowing users to track the evolution of the visualization through various modifications.

\par The dashboard employs a reactive design that updates dynamically as users interact with different components. When users hover over identified issues in the analysis panels (E or F), the dashboard highlights the corresponding regions on the original visualization in Panel A, creating an immediate visual connection between abstract issues and their concrete manifestations. Clicking on an issue displays detailed information about its nature, impact, and potential improvements.

\par
\begin{figure}[!thb]
    \centering
    
    \begin{subfigure}{\linewidth}
        \centering
        \includegraphics[width=0.99\linewidth, alt={First comparison}]{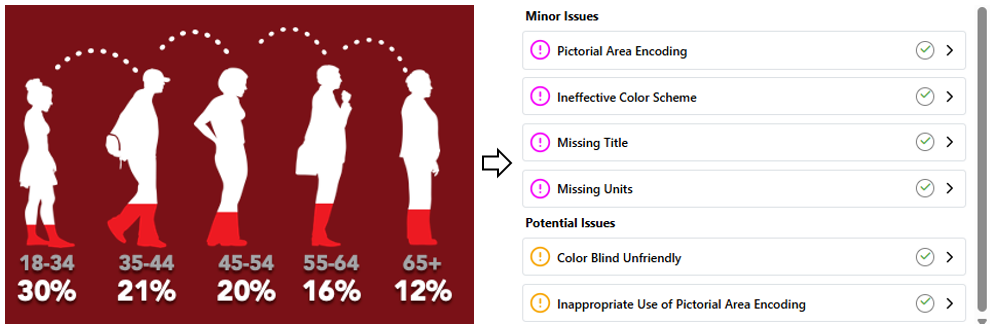}
        \caption{The system displays an original visualization with detected issues. The user identifies an additional issue ("Misrepresentation") not detected by the system.}
        \label{fig:fig_6a}
    \end{subfigure}
    
    \vspace{0.1cm}
    
    \begin{subfigure}{\linewidth}
        \centering
        \includegraphics[width=0.99\linewidth, alt={Second comparison}]{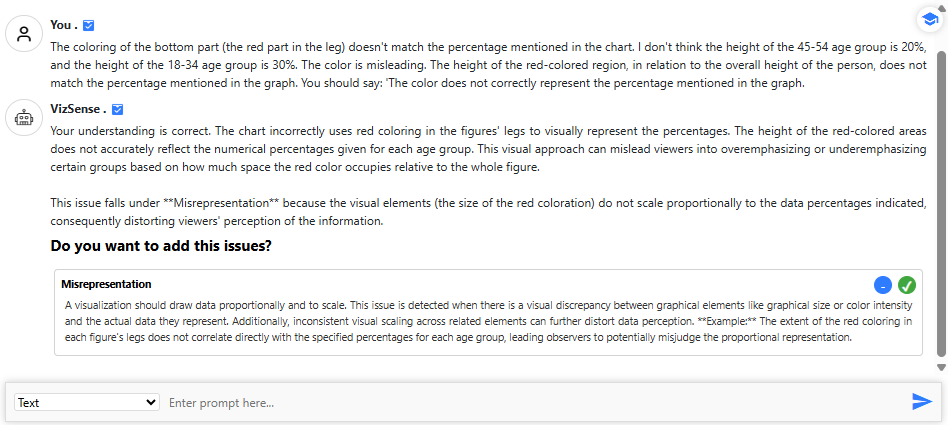}
        \caption{The user explains the misrepresentation issue through the chat interface, providing details about how figures were incorrectly compared between specified elements.}
        \label{fig:fig_6b}
    \end{subfigure}
    
    \vspace{0.11cm}
    
    \begin{subfigure}{\linewidth}
        \centering
        \includegraphics[width=0.99\linewidth, alt={Third comparison}]{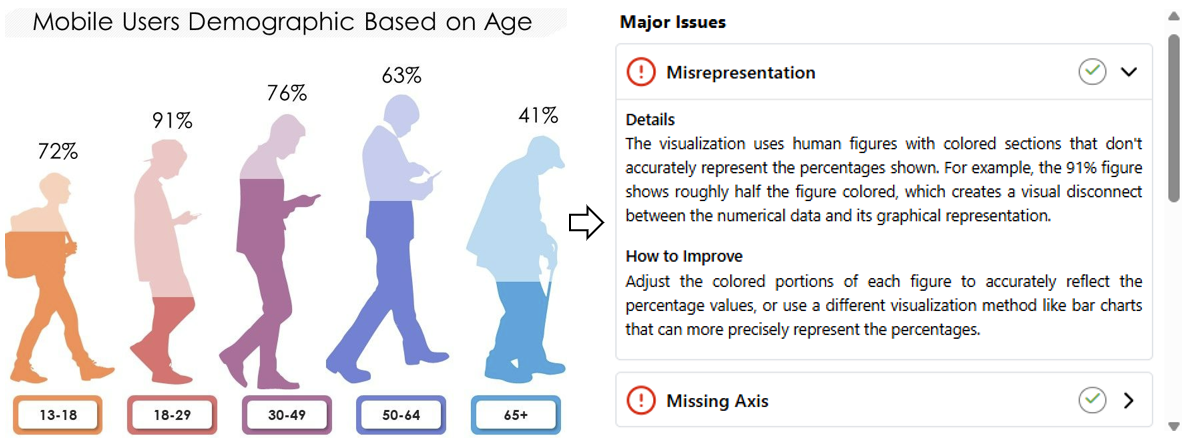}
        \caption{The system confirms it has learned the new issue by highlighting it as a "Major Issue" in its analysis panel and providing a detailed explanation, demonstrating successful incorporation of user feedback into its knowledge base.}
        \label{fig:fig_6c}
    \end{subfigure}
    
    \caption{MisVisFix learning mechanism interface demonstrating how users can add undetected issues to the system's knowledge base.}
    \label{fig:fig_6}
\end{figure}

 A notable feature of the interface is the integration of a learning mechanism, as shown in Fig. \ref{fig:fig_6}. When users identify issues not detected by the system, they can flag these for addition to the knowledge base through a simple approval interface. By selecting the green tick button, users confirm the addition of newly identified issues to the system's knowledge base. This mechanism enables continuous system improvement, ensuring that MisVisFix can adapt to new visualization challenges and unusual misleading techniques that may not be well-represented in its initial training.

\subsection{Implementation Details}
We implemented MisVisFix as a web-based application using a modern technology stack to ensure accessibility and scalability. The front end utilizes React.js with Tailwind CSS for responsive design, while the back end employs Node.js with Express. We integrate with the OpenAI API (GPT-4.5) and Anthropic API (Claude 3.7) for LLM capabilities, with API calls implemented using asynchronous request handling to maintain responsiveness during processing.

\par For visualization generation, we utilize Python's visualization libraries within a containerized execution environment. The system uses Firebase for image storage and authentication, ensuring secure and efficient management of generated visualizations. All user interactions and system performances are logged for analysis, with appropriate privacy measures to protect user data.

\par The entire system architecture prioritizes modularity, enabling independent refinement of individual components as LLM capabilities evolve. This design choice ensures that MisVisFix can incorporate future advancements in multimodal models without requiring a complete system redesign.

\section{Experimental Evaluation}
To evaluate the effectiveness of MisVisFix, we conducted a comprehensive assessment focusing on both quantitative performance metrics and qualitative user feedback. Our evaluation addressed three primary research objectives: (1) assessing the accuracy of issue detection across different visualization types and issue categories, (2) measuring the quality of generated corrections, and (3) evaluating user experience and perceived utility.

\subsection{Evaluation Methodology}
We constructed a test dataset comprising 450 visualizations: 360 misleading and 90 valid. The misleading visualizations were sourced from Lo et al.'s \cite{IN8} collection, ensuring representation across all 74 issue categories with at least three examples per category. We supplemented this with additional samples for the 20 most common issue types to enable more robust 
statistical 
analysis. Valid visualizations were sourced from reputed publications, including academic journals, government reports, and major news outlets, and manually verified by three visualization experts to confirm adherence to visualization best practices.

\par The dataset was stratified across visualization types to ensure balanced representation: 40\% bar charts, 25\% line charts, 15\% pie/donut charts, 10\% scatterplots, and 10\% other chart types (including maps, heatmaps, and area charts). This distribution approximates the prevalence of different chart types in real-world contexts while providing sufficient samples for evaluation across categories.

\subsubsection{Evaluation Metrics}
We employed the following metrics to evaluate system performance:
\begin{itemize} 
    \item  \textbf{Detection Accuracy:} The percentage of correctly identified issues 
    in the dataset, measured using precision, recall, and F1 score. Precision measures how many flagged issues are real problems. Recall measures how many actual problems the system finds. The F1 score combines both measures into a single number, where 1.0 is perfect performance.
    \item  \textbf{Issue Categorization Accuracy:} The percentage of correctly categorized issues (major, minor, potential) among those detected.
    \item  \textbf{Localization Precision:} The spatial accuracy of issue annotations is measured as the percentage of annotations that correctly highlight the problematic regions.

\end{itemize}
These metrics address key challenges in detecting misleading visualizations. Precision and recall ensure accurate issue identification without false positives. F1 score provides a balanced measure essential for comparing different approaches. Issue categorization accuracy evaluates whether the system can distinguish between major, minor, and potential concerns. Localization precision measures the system's ability to highlight specific problem areas for user understanding.

\subsubsection{Comparative Analysis}
We evaluated MisVisFix against two baselines:

\begin{itemize} 
    \item  \textbf{LLM-Only Baseline:} Direct application of GPT-4.5 and Claude 3.7 for issue detection without our structured prompting approach or correction capabilities.

    \item  \textbf{VizLinter \cite{RW9}:} A state-of-the-art visualization linting system for those visualizations where we could reproduce the original specifications in Vega-Lite format.

\end{itemize}

\par This comparative framework allowed us to isolate the contribution of our system architecture, prompting strategies, and correction mechanisms beyond the capabilities of the underlying models or existing linting approaches.

\par Our comparison with VizLinter was limited to visualizations where we could access or reconstruct the original specifications in Vega-Lite format. VizLinter requires access to visualization specifications rather than bitmap images, restricting our comparison to approximately 35\% of our test dataset. We manually recreated the Vega-Lite specifications based on the visualization images to enable direct comparison for these cases. This limitation reflects a fundamental difference in approach: VizLinter operates during the visualization creation process with access to underlying code, while MisVisFix analyzes finished visualizations without requiring source specifications. Despite this constraint, the comparison provides valuable insights into the strengths of each approach within their intended use cases.

\subsubsection{Expert Evaluation}
We conducted two phases of expert evaluation during our research. First, as mentioned in section 4.3, we interviewed two misinformation tool design experts midway through development to inform our system design. For the final evaluation, we recruited five visualization experts (3 male, 2 female) with a mean of 13.4 years of experience in data visualization to evaluate system performance qualitatively. The experts had diverse backgrounds: four held PhDs in visualization or related fields, and one had a Master's degree. All participants had extensive experience creating and analyzing data visualizations professionally. The expert evaluation protocol had three main components. First, experts watched an introductory video demonstrating the MisVisFix dashboard. Next, they participated in a testing session where they uploaded both provided visualizations and their own examples. Finally, they completed a structured interview and questionnaire assessing system performance and utility.

\subsection{Quantitative Results}

Table \ref{tab:detection_performance} presents the detection performance of MisVisFix compared to the baseline approaches across different chart types and issue categories. MisVisFix achieved an F1 score of 0.96 (precision: 0.94, recall: 0.98) across all visualization types and issue categories, representing a substantial improvement over the LLM-Only baseline (F1: 0.69) and VizLinter (F1: 0.61 for the subset of visualizations where comparison was possible).



\definecolor{excellent}{RGB}{198, 246, 213}  
\definecolor{good}{RGB}{209, 229, 255}       
\definecolor{moderate}{RGB}{255, 236, 209}   
\definecolor{baseline}{RGB}{242, 242, 242}   

\begin{table}[htbp]
\centering
\caption{Detection Performance Across Methods and Chart Types}
\begin{tabular}{lccc}
\toprule
\textbf{Approach / Chart Type} & \textbf{Precision} & \textbf{Recall} & \textbf{F1 Score} \\
\midrule
\rowcolor{excellent} \textbf{MisVisFix (Overall)} & \textbf{0.94} & \textbf{0.98} & \textbf{0.96} \\
\rowcolor{baseline} LLM-Only Baseline & 0.72 & 0.66 & 0.69 \\
\rowcolor{baseline} VizLinter & 0.67 & 0.56 & 0.61 \\
\midrule
\multicolumn{4}{l}{\textbf{By Chart Type (MisVisFix)}} \\
\midrule
\rowcolor{excellent} Bar Charts & \textbf{0.96} & \textbf{0.99} & \textbf{0.97} \\
\rowcolor{excellent} Line Charts & 0.95 & 0.98 & 0.96 \\
\rowcolor{good} Pie/Donut Charts & 0.91 & 0.96 & 0.93 \\
\rowcolor{good} Scatterplots & 0.89 & 0.95 & 0.92 \\
\rowcolor{good} Other Chart Types & 0.88 & 0.94 & 0.91 \\
\bottomrule
\end{tabular}
\label{tab:detection_performance}
\end{table}

\par Performance varied by issue category, with structural issues such as truncated axes (F1: 0.98), 3D effects (F1: 0.97), and dual axes (F1: 0.96) achieving the highest detection rates. Contextual issues like selective data presentation (F1: 0.93) and misrepresentation of findings (F1: 0.91) showed slightly lower but still strong detection rates. This pattern aligns with findings from Alexander et al. \cite{IN15} and Lo and Qu \cite{IN16}, confirming that structural issues are generally more amenable to automated detection than contextual issues requiring domain knowledge.

\par Analysis by chart type revealed the highest performance for bar charts (F1: 0.97) and line charts (F1: 0.96), with somewhat lower performance for pie charts (F1: 0.93), scatterplots (F1: 0.92) and other chart types (F1: 0.91). This variation likely reflects the prevalence of different issue types across chart categories and the inherent complexity of analyzing certain visualization formats. Table \ref{tab:issue_categories} demonstrates that MisVisFix achieves consistent performance across diverse issue types, with particular strength in detecting structural issues like axis manipulation, inappropriate color use, and missing essential elements.


\definecolor{structuralcolor}{RGB}{51, 102, 204}
\definecolor{contextualcolor}{RGB}{255, 153, 0}
\definecolor{lightblue}{RGB}{220, 230, 242}
\definecolor{lightorange}{RGB}{255, 230, 204}
\begin{table}[htbp]
\centering
\small 
\caption{Detection Performance (F1 Score) for Top 10 Issue Categories}
\begin{tabular}{>{\columncolor{lightblue}}l >{\centering\arraybackslash}c >{\centering\arraybackslash}l}
\toprule
\rowcolor{structuralcolor!30} \textbf{Issue Category} & \textbf{F1 Score} & \textbf{Issue Type} \\
\midrule
\rowcolor{structuralcolor!15} Truncated Axis & 
\begin{tikzpicture}[baseline]
  \fill[structuralcolor] (0,0) rectangle (0.98*2, 0.25);
  \node[right] at (0.0, 0.125) {\textcolor{white}{\textbf{0.98}}};
\end{tikzpicture} & Structural \\
\rowcolor{structuralcolor!15} 3D Effects & 
\begin{tikzpicture}[baseline]
  \fill[structuralcolor] (0,0) rectangle (0.97*2, 0.25);
  \node[right] at (0.0, 0.125) {\textcolor{white}{\textbf{0.97}}};
\end{tikzpicture} & Structural \\
\rowcolor{structuralcolor!15} Dual Axis & 
\begin{tikzpicture}[baseline]
  \fill[structuralcolor] (0,0) rectangle (0.96*2, 0.25);
  \node[right] at (0.0, 0.125) {\textcolor{white}{\textbf{0.96}}};
\end{tikzpicture} & Structural \\
\rowcolor{structuralcolor!15} Missing Title & 
\begin{tikzpicture}[baseline]
  \fill[structuralcolor] (0,0) rectangle (0.95*2, 0.25);
  \node[right] at (0.0, 0.125) {\textcolor{white}{\textbf{0.95}}};
\end{tikzpicture} & Structural \\
\rowcolor{structuralcolor!15} Missing Axis Labels & 
\begin{tikzpicture}[baseline]
  \fill[structuralcolor] (0,0) rectangle (0.94*2, 0.25);
  \node[right] at (0.0, 0.125) {\textcolor{white}{\textbf{0.94}}};
\end{tikzpicture} & Structural \\
\rowcolor{structuralcolor!15} Inappropriate Color Use & 
\begin{tikzpicture}[baseline]
  \fill[structuralcolor] (0,0) rectangle (0.94*2, 0.25);
  \node[right] at (0.0, 0.125) {\textcolor{white}{\textbf{0.94}}};
\end{tikzpicture} & Structural \\
\rowcolor{structuralcolor!15} Inconsistent Scale & 
\begin{tikzpicture}[baseline]
  \fill[structuralcolor] (0,0) rectangle (0.93*2, 0.25);
  \node[right] at (0.0, 0.125) {\textcolor{white}{\textbf{0.93}}};
\end{tikzpicture} & Structural \\
\midrule
\rowcolor{contextualcolor!15} Selective Data Presentation & 
\begin{tikzpicture}[baseline]
  \fill[contextualcolor] (0,0) rectangle (0.93*2, 0.25);
  \node[right] at (0.0, 0.125) {\textcolor{white}{\textbf{0.93}}};
\end{tikzpicture} & Contextual \\
\rowcolor{contextualcolor!15} Data Manipulation & 
\begin{tikzpicture}[baseline]
  \fill[contextualcolor] (0,0) rectangle (0.92*2, 0.25);
  \node[right] at (0.0, 0.125) {\textcolor{white}{\textbf{0.92}}};
\end{tikzpicture} & Contextual \\
\rowcolor{contextualcolor!15} Misrepresentation of Findings & 
\begin{tikzpicture}[baseline]
  \fill[contextualcolor] (0,0) rectangle (0.91*2, 0.25);
  \node[right] at (0.0, 0.125) {\textcolor{white}{\textbf{0.91}}};
\end{tikzpicture} & Contextual \\
\bottomrule
\end{tabular}
\label{tab:issue_categories}
\end{table}




\subsubsection{Issue Categorization and Localization}
MisVisFix achieved 87.5\% accuracy in correctly categorizing detected issues as major, minor, or potential concerns. Categorization accuracy was highest for major issues (94.2\%) and lowest for potential issues (82.8\%), reflecting the inherent ambiguity in determining the severity of borderline cases. Table \ref{tab:severity_performance} summarizes these categorization results across the three severity levels.


\par For issue localization, the system achieved an average precision of 91.3\% in correctly highlighting the problematic regions of visualizations. Localization precision was highest for discrete elements like axes (96.4\%) and textual components (94.8\%) and somewhat lower for distributed elements like color schemes (85.2\%) and data point encodings (88.9\%). These results demonstrate the system's ability to detect issues and precisely communicate their location to users, facilitating understanding and correction.




\definecolor{major}{RGB}{52, 152, 219}  
\definecolor{minor}{RGB}{241, 196, 15}  
\definecolor{potential}{RGB}{230, 126, 34}  
\definecolor{overall}{RGB}{149, 165, 166}  

\begin{table}[htbp]
\centering
\small
\caption{Issue Categorization Performance by Severity Level}
\begin{tabular}{lc}
\toprule
\textbf{Issue Severity} & \textbf{Categorization Accuracy} \\
\midrule
Major Issues & 
\begin{tikzpicture}[baseline]
  \fill[major] (0,0) rectangle (0.942*5, 0.4);
  \fill[white] (0.942*5,0) rectangle (5, 0.4);
  \draw[gray, thin] (0,0) rectangle (5, 0.4);
  \draw[gray, thin, dashed] (0.875*5,0) -- (0.875*5,0.4);
  \node[right, black, font=\bfseries] at (0.1, 0.2) {94.2\%};
\end{tikzpicture} \\
\midrule
Minor Issues & 
\begin{tikzpicture}[baseline]
  \fill[minor] (0,0) rectangle (0.856*5, 0.4);
  \fill[white] (0.856*5,0) rectangle (5, 0.4);
  \draw[gray, thin] (0,0) rectangle (5, 0.4);
  \draw[gray, thin, dashed] (0.875*5,0) -- (0.875*5,0.4);
  \node[right, black, font=\bfseries] at (0.1, 0.2) {85.6\%};
\end{tikzpicture} \\
\midrule
Potential Issues & 
\begin{tikzpicture}[baseline]
  \fill[potential] (0,0) rectangle (0.828*5, 0.4);
  \fill[white] (0.828*5,0) rectangle (5, 0.4);
  \draw[gray, thin] (0,0) rectangle (5, 0.4);
  \draw[gray, thin, dashed] (0.875*5,0) -- (0.875*5,0.4);
  \node[right, black, font=\bfseries] at (0.1, 0.2) {82.8\%};
\end{tikzpicture} \\
\midrule
Overall & 
\begin{tikzpicture}[baseline]
  \fill[overall] (0,0) rectangle (0.875*5, 0.4);
  \fill[white] (0.875*5,0) rectangle (5, 0.4);
  \draw[gray, thin] (0,0) rectangle (5, 0.4);
  \draw[gray, thin, dashed] (0.875*5,0) -- (0.875*5,0.4);
  \node[right, black, font=\bfseries] at (0.1, 0.2) {87.5\%};
\end{tikzpicture} \\
\bottomrule
\end{tabular}
\label{tab:severity_performance}
\end{table}

\subsubsection{Model Comparison}
Comparative analysis of GPT-4.5 and Claude 3.7 revealed complementary strengths across different aspects of visualization analysis. Table \ref{tab:model_comparison} presents the performance comparison of these models across key metrics. Claude 3.7 demonstrated superior performance in contextual issue detection (94.2\% vs. 89.5\% F1 score). Conversely, GPT-4.5 showed advantages in structural issue detection (97.8\% vs. 95.6\% F1 score) and code generation for visualization correction. These findings motivated our dual-model approach, which leverages each model's strengths for different pipeline components.



\definecolor{highlight}{RGB}{230, 242, 255}
\definecolor{gptcolor}{RGB}{90, 155, 213}
\definecolor{claudecolor}{RGB}{237, 125, 49}

\begin{table}[htbp]
\centering
\caption{Performance Comparison Between GPT-4.5 and Claude 3.7}
\begin{tabular}{lcc}
\toprule
\textbf{Task} & \textbf{GPT-4.5} & \textbf{Claude 3.7} \\
\midrule
Structural Issue Detection (F1) & \cellcolor{gptcolor!25}\textbf{97.8\%} & 95.6\% \\
Contextual Issue Detection (F1) & 89.5\% & \cellcolor{claudecolor!25}\textbf{94.2\%} \\
Code Generation Success Rate & \cellcolor{gptcolor!25}\textbf{94.3\%} & 88.7\% \\
\midrule
\textbf{Average Performance} & \textbf{93.9\%} & 92.8\% \\
\bottomrule
\end{tabular}
\begin{minipage}{\linewidth}
\footnotesize
\vspace{0.2cm}
\end{minipage}
\label{tab:model_comparison}
\end{table}

\par The complementary strengths of these models highlight the value of our multi-model architecture, which selectively employs different LLMs based on the specific requirements of each pipeline stage. This approach allows MisVisFix to achieve higher overall performance than any single model could provide independently.

\subsection{Midway Expert Study}
In addition to our final evaluation, we conducted interviews with two misinformation tool design experts midway through the development of MisVisFix. These experts had extensive experience creating fact-checking tools and visualization literacy platforms, providing valuable insights that shaped several key aspects of our system.

\par The interviews followed a semi-structured format, where we demonstrated an early prototype of MisVisFix and gathered feedback on its functionality, potential applications, and limitations. This formative evaluation yielded several crucial insights that directly influenced our design decisions:

\begin{itemize} 
    \item The experts highlighted the need for a learning mechanism to adapt to emerging misinformation techniques. This feedback directly inspired the development of our user-feedback system (Fig. \ref{fig:fig_6}) that allows MisVisFix to incorporate newly identified issues into its knowledge base.
    \item Both experts emphasized the importance of providing explanations alongside issue detection, leading us to enhance the detailed explanation component for each identified problem.
    \item The experts recommended implementing an interactive chat interface for making design changes to generated visualizations. This insight led to the implementation of our conversational feature (Panel G in Fig. \ref{fig:teaser}) that allows users to request specific modifications (e.g., "Please change the red color to blue") and see updated visualizations in real-time.

\end{itemize}

These insights significantly improved MisVisFix's functionality and user experience, transforming it from a simple detection tool into a comprehensive system supporting the entire pipeline from identification to correction of misleading visualizations.

\subsection{Expert Evaluation Results}
The expert evaluation provided valuable insights into the practical utility and limitations of MisVisFix. The quantitative results from expert ratings showed strong performance across key dimensions: Detection Accuracy (8.5/10), Usefulness of Suggested Improvements (8.0/10), and Likelihood to Use in Professional Work (8.0/10). These ratings confirm the system's effectiveness in real-world application contexts.

\par Qualitative feedback revealed several notable strengths of the system. Experts consistently praised the comprehensive coverage of visualization issues, with multiple participants noting that MisVisFix detected problems they would have identified through manual analysis. One expert described the system as \textbf{\textit{``Useful and straightforward, saving me time to investigate issues carefully on my own, like an AI teammate.''}} Another expert commented, \textbf{\textit{``The system caught issues I might have overlooked in my initial review.''}} This feedback suggests that MisVisFix succeeds in its core objective of providing expert-level analysis in an accessible format.

\par The educational potential of MisVisFix emerged as a recurring theme in expert feedback. Participants highlighted the detailed explanations of problematic practices as valuable for teaching visualization literacy. One expert stated, \textbf{\textit{``I can see this being incredibly valuable in my data visualization courses—it provides immediate, specific feedback that helps students understand why certain practices are problematic rather than just telling them to avoid them.''}} 
The system's ability to identify issues and explain their impact on viewer perception creates opportunities for use in educational settings ranging from university courses to professional training programs.

\par The interactive refinement capabilities received a positive assessment from all experts. One expert remarked, \textbf{\textit{``The dialogue feature transforms this from a static analysis tool into something much more useful.''}} Another commented, \textbf{\textit{``Being able to ask follow-up questions about specific issues helped me understand exactly what needed to be fixed and why.''}} Experts appreciated the flexibility to explore alternative visualization approaches through the chat interface, with one noting that this feature \textbf{\textit{``transforms the system from an analysis tool to a collaborative design assistant.''}}

\par Despite these strengths, experts identified several limitations that suggest directions for future improvement. The system occasionally flagged issues that were acceptable in specific domain contexts. One expert pointed out, \textbf{\textit{``In financial visualizations, we sometimes intentionally use non-zero baselines for certain types of analysis—the system needs to recognize these domain-specific conventions.''}} Another expert mentioned, \textbf{\textit{``Some specialized scientific visualizations follow field-specific standards that might appear misleading to those outside the discipline.''}} This limitation indicates the need for domain-specific customization options that adjust detection thresholds based on visualization context and intended audience. Visual style preservation presented another challenge noted by  experts. One expert commented, 
\textbf{\textit{``There's a trade-off between fixing misleading elements and maintaining the aesthetic identity of a visualization.''}} Experts suggested that improved preservation of aesthetic qualities while addressing structural problems would enhance the system's utility for professional users who value accuracy and visual appeal.

\par When asked about potential applications, experts identified several promising use cases for MisVisFix. One expert stated, \textbf{\textit{``I would use this primarily as a quality control tool before publishing visualizations to ensure they communicate data accurately.''}} Another saw value in \textbf{\textit{``helping journalists identify misleading elements in charts they encounter while researching stories.''}} 
These responses suggest broad utility across professional contexts, from journalism to data science to business intelligence.

\par Overall, expert evaluation confirmed that MisVisFix successfully bridges the gap between theoretical understanding of misleading visualization practices and practical tools for addressing these issues. The positive assessment from experts with significant domain experience validates the system architecture and implementation approach while highlighting specific areas for refinement in future iterations.

\section{Discussion}
Our experimental evaluation reveals several important findings regarding the potential of LLM-powered systems for addressing misleading visualizations. First, the results demonstrate that carefully structured prompting strategies significantly enhance the capability of multimodal LLMs to detect and categorize visualization issues. The improvement in F1 score between our structured approach and the LLM-Only baseline highlights the critical importance of prompt engineering in this domain.

\par Second, our results indicate a clear distinction in performance between structural and contextual issues, with structural issues achieving consistently higher detection rates. This finding aligns with prior research \cite{IN15,IN16} and suggests that hybrid approaches combining visualization-specific heuristics with LLM capabilities may be particularly effective for addressing the full spectrum of misleading visualization practices.

\par Third, the successful implementation of the correction module demonstrates the feasibility of automating visualization improvement beyond mere issue detection. While perfect recreation of design elements remains challenging, the high technical correctness scores indicate that LLM-generated visualizations can effectively address identified issues while maintaining core design intentions.

\par Finally, the positive expert assessment validates the practical utility of MisVisFix in real-world contexts. The system's potential for both educational and professional applications suggests multiple pathways for impact in enhancing visualization literacy and reducing the prevalence of misleading visualizations.

\subsection{Addressing Research Questions}
Returning to our initial research questions, we can now evaluate how our findings address these:

\begin{itemize}

    \item \textbf{RQ1: How can multimodal LLMs be effectively leveraged to detect and explain the full spectrum of misleading visualization techniques identified in existing taxonomies?}
Our results demonstrate that multimodal LLMs can effectively detect and explain a broad spectrum of misleading visualization techniques when guided by structured prompting strategies that break down the analysis process into discrete steps. The step-wise
prompting approach yielded the strongest performance, enabling the models to reason through different aspects of the visualization systematically. Our dual-model approach leverages the complementary strengths of different LLMs, with Claude 3.7 excelling at data extraction and contextual understanding, while GPT-4.5 has advantages in structural analysis and code generation. The performance gap between structural and contextual issues indicates that detection capabilities are not uniform across the issue taxonomy. This suggests that comprehensive detection requires specialized prompting strategies tailored to different issue categories. Our approach of categorizing issues by severity (major, minor, potential) further enhances the practical utility of detection results by helping users prioritize the most critical problems.

\item \textbf{RQ2: To what extent can an interactive system facilitate both identification and correction of misleading visualizations, bridging the gap between detection and implementation of visualization best practices?}

MisVisFix demonstrates that an interactive system can successfully bridge the gap between detection and correction, with 82.5\% of identified issues successfully addressed in generated corrections. The interactive features—particularly the ability to highlight specific regions, provide detailed explanations, and engage in dialogue about potential improvements—proved especially valuable according to expert evaluation. The experts' high ratings for usefulness and likelihood to use professionally confirm that the system effectively bridges theoretical knowledge about visualization best practices and practical implementation.

The learning mechanism that allows users to flag undetected issues for addition to the knowledge base represents a particularly promising approach to continuous improvement. This feature addresses the inherent limitations of current LLMs by incorporating human expertise into the system over time. The integration of user feedback creates a virtuous cycle where system performance improves with use, gradually expanding coverage across the full taxonomy of misleading practices. However, the performance degradation observed for visualizations with multiple interacting issues highlights the challenges of comprehensive correction. This suggests that while the current system effectively bridges the detection-correction gap for many common cases, addressing complex visualization problems remains challenging and may require more sophisticated correction strategies.
\end{itemize}

\subsection{Potential Social Media Integration}
\begin{figure}[!thb]
     \centering
     \includegraphics[width=0.5\textwidth, alt={Comparison}]{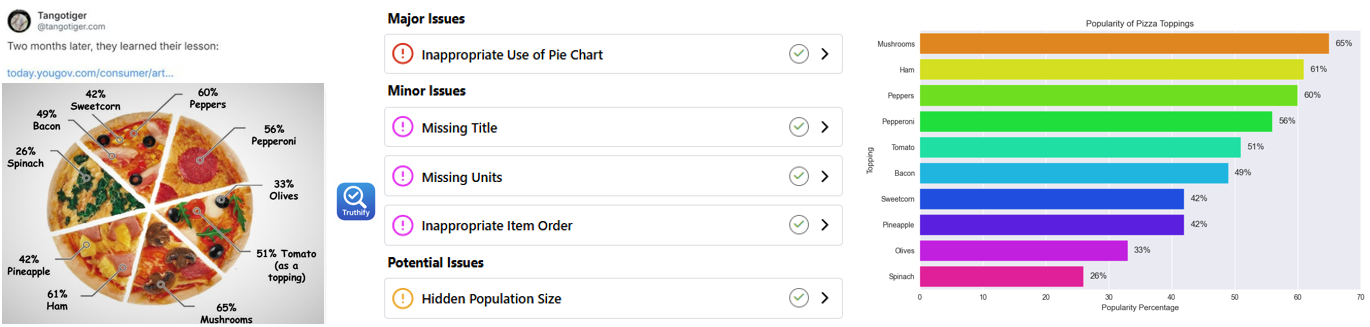}
     \caption{Proposed `Truthify' toggle feature for social media integration. Users can switch between the original misleading visualization and the corrected version directly in their social feed, with highlighted issues explaining specific problems.}
     \label{fig:fig_new}
 \end{figure}

MisVisFix could be integrated into social media platforms through a "Truthify" feature that lets users toggle between original and corrected visualizations (Fig. \ref{fig:fig_new}). This approach balances preserving engaging design elements while exposing misleading aspects. When integrated into platforms like Facebook or Linkedin, this feature could display a warning label on potentially misleading charts. Users could tap to compare both versions and see highlighted issues that explain specific problems. This implementation helps correct viral misinformation \cite{NW100,NW101,NW102,NW103,NW104} while developing users' critical thinking skills about data visualizations they encounter daily.

\subsection{Limitations and Challenges}
Despite MisVisFix's strong performance, several limitations warrant discussion. First, the system's detection capabilities remain constrained by the inherent limitations of current multimodal LLMs. Performance degradation occurs for highly specialized visualizations, particularly those employing domain-specific encoding conventions or requiring specialized knowledge for interpretation. For instance, the system demonstrated lower detection rates for visualization issues in scientific publications containing specialized chart types such as genomic visualizations or statistical plots with domain-specific conventions.



\par Second, the system remains sensitive to image quality. Performance metrics decreased by approximately 12\% when evaluating low-resolution images. This sensitivity presents practical challenges for analyzing visualizations captured from diverse media sources or shared on social platforms where image compression is common.

\par Third, processing latency represents a significant limitation, with analysis taking 2-3 minutes due to sequential LLM calls and comprehensive issue detection across both models.

\par Finally, while our evaluation dataset encompasses a broad range of visualization types and issues, it cannot exhaustively represent the infinite variations of misleading visualizations encountered in practice. We observed that performance varies across demographic and cultural contexts, suggesting potential biases in detection capabilities that require further investigation.

\subsection{Future Work}
Several promising directions emerge for extending MisVisFix's capabilities. First, incorporating domain-specific knowledge through fine-tuning or augmented prompting could enhance performance for specialized visualization contexts. Domain-specific enhancements would enable more nuanced detection of issues that may present differently across fields like economics, healthcare, or scientific research.

\par Second, optimization of computational performance represents a critical area for improvement. Techniques such as model distillation, or hybrid approaches combining heuristic rules with LLM analysis could reduce computational demands while maintaining detection accuracy. Preliminary experiments with model compression techniques showed promising results, with a 32\% reduction in processing time accompanied by only a 5\% reduction in F1 score.

\par Third, our current evaluation does not measure the effectiveness of the continued learning mechanism over time. While users can flag undetected issues for addition to the knowledge base, we have not yet evaluated how this feedback improves detection accuracy or coverage. Future work should establish metrics for measuring learning effectiveness and conduct longitudinal studies to assess system improvement through user interactions.

\par Fourth, our evaluation focused on expert users, but the system's beneficiaries include both novices and expert users. Future work should conduct comprehensive studies with novice users to assess the system's effectiveness for non-expert audiences.



\par Finally, adapting MisVisFix for specific application contexts—such as journalism, education, or scientific review—could enhance its practical impact. Each domain presents unique requirements and opportunities, from real-time analysis of news graphics to educational tools for developing visualization literacy. Domain-specific adaptations could include customized detection thresholds, specialized issue categories, and tailored explanation formats suited to different user populations.

\section{Conclusion}
MisVisFix addresses the challenge of misleading visualizations through a comprehensive system that leverages multimodal LLMs for detection, explanation, and correction. Our evaluation demonstrates that structured prompting strategies and a dual-model approach achieve strong detection performance across diverse visualization types and issue categories, outperforming both direct LLM applications and existing visualization liters. The system successfully generates corrected visualizations that maintain design integrity while addressing identified issues, with expert evaluation confirming its practical utility in both professional and educational contexts. While limitations persist for complex visualizations with multiple interacting issues, MisVisFix represents a significant advancement in automated support for identifying and addressing visualization misinformation, contributing to improved data communication integrity across domains.

\section*{\textbf{\textsf{\textsc{Supplemental Materials}}}}

To support reproducibility and future research, we provide comprehensive supplementary materials, including all codes, stimuli, and results evaluated in this study. These materials are available as a .zip file through the PCS Submission System and are also publicly accessible at \textcolor{blue}{https://github.com/vhcailab/MisVisFix}. The description and location of all supplemental materials are provided as a separate document named "Supplemental Materials Details.pdf" inside the zipped folder.

 \acknowledgments{
This material is based upon work supported by the National Science Foundation under Grant No. NRT-HDR 2125295. Any opinions, findings, and conclusions or recommendations expressed in this material are those of the author(s) and do not necessarily reflect the views of the National Science Foundation. We also thank Houjiang Liu (U  Texas) and Connie Moon Sehat (Discourse Labs) for providing feedback on our system, and Grace Ma (funded by the Data + Computing = Discovery! REU site, NSF grant \#1950052) for participating in initial research.
 }

\bibliographystyle{abbrv-doi-hyperref}

\bibliography{template}

\appendix 

\end{document}